# Smart transformer Modelling in Optimal Power Flow Analysis

Junru Chen, Ran Li, Alireza Soroudi, Andrew Keane, Damian Flynn, Terence O'Donnell
University College Dublin
Dublin, Ireland
junru.chen.1; ran.li@ucdconnect.ie;
alireza.soroudi; andrew.keane; damian.flynn; terence.odonnell@ucd.ie

*Abstract*— The smart transformer (ST) implemented using power electronics converters, has the capability of independent voltage control and reactive power isolation between it primary and secondary terminals. This capability provides a flexibility in the power system to support the voltage at the primary side and control the demand at the secondary side. Using this flexibility, the system power flow could for example be optimized for lower costs. This paper proposes an ST model suitable for OPF analysis. The effects of using multiple STs at different penetration levels, on the daily generation costs in an IEEE 39 bus test system are presented.

*Keywords—Smart Transformer, Static Model, Economical Dispatch, Optimal Power Flow*

## I. INTRODUCTION

With the continuing increase in the deployment of distributed generation, the control and coordination of the interface between the transmission and distribution systems become critically important. This requires increased flexibility from the distribution system through demand response, frequency support from DER and voltage support through the provision of reactive power. In this context, the smart transformer [1] provides an interesting approach to the provision of such flexibility at the transmission-distribution interface.

The ST is usually implemented as a three-stage solid state transformer [2,3] with the capability for ancillary system service provision. In this topology, in addition to a primary and secondary AC port, the ST can also provide high and low voltage DC ports, with a potential to connect to DC subsystems, such as, renewable generation [4], electric storage [5,6] and electric vehicle charger stations [7]. Since the voltage at each port is controlled by the different stage converters, the voltages for each port can be fully decoupled. Exploiting this feature, the secondary voltage can be used to identify the load voltage sensitivity [8,9], minimize the neutral current arising from unbalanced loads [10] and dynamically regulate the demand [11] in response to the grid frequency [12]. On the other hand, the capacitor in the DC ports helps to isolate the reactive power for each port; thus, the ST only delivers the active power while the reactive power for each port is fully independent [3]. Thus, the ST can be used to compensate the reactive power in the primary side to enhance the voltage stability [13-15].

To date much of the ST research has been dedicated to the ST topology, and the development of additional device level controls, with ancillary service provision being more of a focus in recent years. While such work is important to understand the ST capabilities and possibilities, more system level studies will be vital to understand its role in solving system level problems especially in the context of increased renewable generation. For example, since the primary and secondary voltage and reactive power in the ST are fully decoupled, this gives a flexibility to optimize the power flow in the system. To date the role that the ST might play in optimal power flow has not being widely researched. Reference [16] has proposed the static model of a multi-port ST (called an energy router), and based on this model, [17,18] investigates the benefit of the multi-port ST as a power flow controller using OPF in an IEEE 24-bus system and IEEE 118 bus system. In that work the energy router was envisaged as being embedded in the transmission system. In contrast to that, in this paper we investigate the use of the ST at the interface between the transmission system and distribution system. In this case the primary function of the ST should be supplying the load with the provision of ancillary service such as voltage and frequency support as secondary functions. Hence the control approach and model required is different from that required for the energy router. In order to study the system level benefits of the ST in larger systems a model suitable for use in optimization studies is required. The full-switching EMT ST model is commonly used in smaller system power flow analysis [19], but is inappropriate in larger systems due to the computational burden. Reference [20,21] proposed the differential-algebraic equation (DAE) based model of the ST, taking into account the dynamics from the controller and filter transients. However, for the power flow analysis, it is not necessary to emulate the ST transient response, because the system is stabilized at one operation point. This paper proposes a static model of the ST suitable for use in OPF studies, which represents an ST connected at one side to the transmission system and at the other side to the distribution system loads. The model includes the ST capability to provide reactive power and hence voltage support to the transmission system and to control demand by varying the voltage on the distribution side.

The model is then used in an IEEE 39 bus system as a case study to quantify the benefits of the ST in terms of total generation costs reduction. Moreover the benefits of increasing the number of STs used in the system are also investigated. This is done by varying the number of STs used at the transmission distribution interfaces from initially at just one interface to all such interfaces.

The rest of the paper is organized as follows: Section II reviews the ST structure and develops the ST static model. Section III introduces the OPF formulation in the transmission system. Section IV presents the case study using the IEEE 39-bus system to quantify the generation cost reduction.

## II. SMART TRANSFORMER MODELLING

The paper focuses on the OPF analysis based on the transmission system, thus, the ST consists of the HVAC-HVDC rectifier connecting to the transmission system, HVDC-MVDC DC-DC converter, considered to be implemented as a dual active bridge converter (DAB) and MVDC-MVAC inverter feeding to the loads as shown in Fig. 1. The HVAC-HVDC rectifier synchronizes to the grid via a phase locked loop (PLL) and applies outer power, inner current control to regulate the MVDC voltage. The DAB converter contains a medium frequency transformer and aims to regulate the MVDC voltage. The MVDC-MVAC inverter applies outer voltage, inner current control to regulate the MVAC voltage to the distribution system or loads. In this topology, only the active power flows through each converter, while reactive power is fully decoupled at each port, due to the inclusion of the capacitor in the DC links. In addition, the voltage at each port is also independent, and can be controlled to track its reference. Because of these characteristics, the ST on the transmission system side can compensate reactive power to support the voltage, which is similar to a STATCOM, and on the distribution system side, can provide frequency support through demand control. The details of the dynamics and control for these ST functions are fully described in [12] and the differential-algebraic equation (DAE) model of the ST is proposed in [20,21]. In this paper, we aim to illustrate the benefit of the ST on the economic operation of a power system. To address this problem, a static ST model needs to be considered in system level optimal power flow analysis. The rest of this section introduces the static model of the ST.

The static model only focuses on the static power flow in the system and neglects all the dynamics and harmonics. Thus, the static model can be deduced from the DAE model in [20,21] by the means of forcing the differential equations to be zero, neglecting the controller dynamics and only considering the nominal frequency.

### A. HVAC-HVDC Rectifier Static Model

In the HVAC side, the ST connects to the grid at the point of common coupling bus, which is the bus before the filter of the rectifier. From the transmission point of view, the connected ST is a PQ bus $i$, with active and reactive power $(P_{ST,i}, Q_{ST,i})$. The power flow determines the voltage amplitude and phase $(V_i \angle \delta_i)$. In steady state, the PLL locks the phase so that the active and reactive power can be fully decoupled in synchronous dq-frame at the PCC point as in (1), where $I_{drec,i} + jI_{qrec,i}$ is the current from the grid at bus $i$ flowing into the rectifier.

$$\left. \begin{array}{l} P_{ST,i} = I_{drec,i}V_i \\ Q_{ST,i} = I_{qrec,i}V_i \end{array} \right\} \quad (1)$$

Defining $R_{frec,i} + jX_{frec,i}$ as the impedance of the filter, $m_{drec,i} + jm_{qrec,i}$ is the rectifier modulation index, then the electrical relationships between the HVAC and HVDC side of the rectifier can be formulated as (2) where $V_{DCh,i}$ is the HVDC voltage.

$$\left. \begin{array}{l} m_{drec,i}\dfrac{V_{DCh,i}}{2} = V_i - R_{frec,i}I_{drec,i} - X_{frec,i}I_{qrec,i} \\ m_{qrec,i}\dfrac{V_{DCh,i}}{2} = -R_{frec,i}I_{qrec,i} + X_{frec,i}I_{drec,i} \end{array} \right\} \quad (2)$$

### B. MVDC-LVDC DAB Static Model

In steady state, the DC bus voltages $V_{DCh,i}$ and $V_{DCm,i}$ for both HVDC and MVDC are constant, thus, the capacitor dynamics and control actions can be neglected. The core part here is the active power delivery and associated losses in this process. The active power passes DAB is the power at the PCC point minus the losses in the filter of the rectifier as (3).

$$P_{DAB,i} = P_{ST,i} - R_{frec,i}I_{dinv,i} \quad (3)$$

If it is assumed that $N_{DC,i}$ is the voltage ratio from HVDC to MVDC ($N_{DC,i} = V_{DCh,i}/V_{DCm,i}$), including the combination of the transformer ratio and DAB modulation, and $R_{DCh,i}$ and $R_{DCm,i}$ represent losses on both sides of the DAB, then the DAB losses can be computed as (4).

$$P_{DABl,i} = \dfrac{P_{DAB,i}}{V_{DCm,i}}(R_{DCh,i} + R_{DCm,i}N_{DC,i}^2) \quad (4)$$

where $\dfrac{P_{DAB,i}}{V_{DCh,i}}$ is the DC current, $R_{DCm,i}N_{DC,i}^2$ is the MVDC resistance reflected to the HVDC side.

### C. MVDC-MVAC Inverter Static Model

The inverter in MVAC side connects to the distribution system feeding the load directly. In the static power flow analysis, the loading at the nominal voltage ($V_{s0}$) is assumed to be a given parameter ($P_{L0,i}, Q_{L0,i}$). Of course, the real loading is also related to the supply voltage or the inverter output voltage $V_s$, the relationship for which is commonly modelled as (5).

$$\left. \begin{array}{l} P_{L,i} = P_{L0,i}(\dfrac{V_{s,i}}{V_{s0,i}})^{\alpha_i} \\ Q_{L,i} = Q_{L0,i}(\dfrac{V_{s,i}}{V_{s0,i}})^{\beta_i} \end{array} \right\} \quad (5)$$

where $\alpha_i$ and $\beta_i$ is the load active and reactive voltage sensitivity.

The inverter output voltage at the MVAC side can be fully controlled, thus, it can be set to $V_{s,i} \angle 0$, meaning that the current to power relationship in a synchronous dq-frame is decoupled as (6).

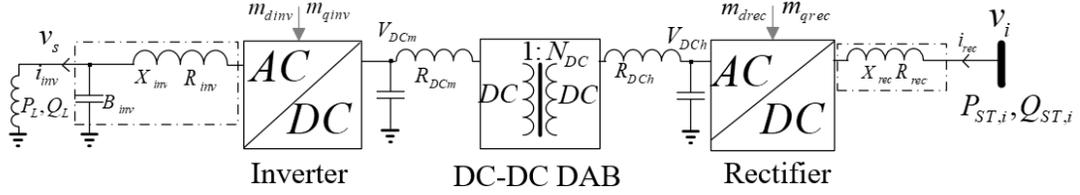

Fig. 1. Smart Transformer Topology

$$I_{dinv,i} = \frac{P_{L,i}}{V_{s,i}} \\ I_{qinv,i} = -\frac{Q_{L,i}}{V_{s,i}}\} \quad (6)$$

Defining $R_{finv,i} + jX_{finv,i} + jB_{finv,i}$ as the impedance and susceptance of the filter, $m_{d,inv} + jm_{q,inv}$ is the inverter modulation index, then the electrical relationships between the MVAC and MVDC side of the inverter can be computed as (7).

$$V_{s,i} = m_{dinv,i}\frac{V_{DCm,i}}{2} - R_{finv,i}I_{dinv,i} - X_{finv,i}(I_{qinv,i} + B_{finv,i}I_{dinv,i}) \\ 0 = m_{qinv,i}\frac{V_{DCm,i}}{2} - R_{finv,i}(I_{qinv,i} + B_{finv,i}I_{dinv,i}) + X_{finv,i}I_{dinv,i}\} \quad (7)$$

The ST only transfers active power, thus, the active power drawn by the ST at the PCC bus on the HV side is given by (8).

$$P_{ST,i} = R_{frec,i}I_{dinv,i} + P_{DABm,i} + P_{L,i} + R_{finv,i}I_{dinv,i} \quad (8)$$

### D. ST Constraints

The EN50160 standard requires that the load supply voltage should be in a range, thus:

$$V_{s,i}^{min} \leq V_{s,i} \leq V_{s,i}^{max} \quad (9)$$

In addition, the ST has strict current limitations for each part, i.e.:

$$0 \leq I_{drec,i}^2 + I_{drec,i}^2 \leq I_{rec,i}^{max\,2} \\ 0 \leq I_{dinv,i}^2 + I_{dinv,i}^2 \leq I_{inv,i}^{max\,2}\} \quad (10)$$

The modulation for the converters should be within ±1, i.e.:

$$\begin{array}{l}-1 \leq m_{drec,i} \leq 1 \\ -1 \leq m_{qrec,i} \leq 1 \\ -1 \leq m_{dinv,i} \leq 1 \\ -1 \leq m_{qinv,i} \leq 1\end{array}\} \quad (10)$$

Equation (1-8) and constraints (9-10) define the static model of the ST at bus *i* used for the OPF analysis. The ST fixed parameters are the load information ($P_{L0,i}$, $Q_{L0,i}, V_{s0,i}, \alpha_i, \beta_i$), DC voltages ($V_{DCh,i}, V_{DCm,i}$) and ST component values ($R_{frec,i}, X_{frec,i}, R_{DCh,i}, R_{DCm,i}, N_{DC,i}, R_{finv,i}, X_{finv,i}, B_{finv,i}$). The variables are PCC power ($P_{ST,i}, Q_{ST,i}$), the currents ($I_{drec,i}, I_{qrec,i}, I_{dinv,i}, I_{qinv,i}$), the modulation indices ($m_{drec,i}, m_{qrec,i}, m_{dinv,i}, m_{qinv,i}$) and LVAC voltage ($V_{s,i}$). Here we define the matrix for the variables as $X_{ST}$ (11), and the constraints as $X_{ST}^{max}$(12), $X_{ST}^{min}$ (13).

$$X_{ST}^{max} = [V_{s,i}^{max}, I_{rec,i}^{max}, I_{inv,i}^{max}, 1, 1, 1, 1]^T \quad (12)$$

$$X_{ST}^{min} = [V_{s,i}^{min}, 0, 0, 1, 1, 1, 1]^T \quad (13)$$

## III. OPF FORMULATION FOR THE TRANSMISSION SYSTEM

Those ST's variables related to the OPF analysis are the power at the PCC point ($P_{ST,i}, Q_{ST,i}$), which is a PQ bus *i* in the transmission system. The OPF formulation has been well developed in previous literature [22], and this section gives a very brief review to illustrate the equations.

### A. Optimal AC Power Flow

The power flow ($P_{ij}, Q_{ij}, S_{ij}$) between two buses is computed as (14).

$$P_{ij} = \frac{V_i^2}{Z_{ij}}cos(\theta_{ij}) - \frac{V_iV_j}{Z_{ij}}cos(\delta_i - \delta_j + \theta_{ij}) \\ Q_{ij} = \frac{V_i^2}{Z_{ij}}sin(\theta_{ij}) - \frac{V_iV_j}{Z_{ij}}sin(\delta_i - \delta_j + \theta_{ij}) - \frac{b_{ij}V_i^2}{2} \\ |S_{ij}| = \sqrt{P_{ij}^2 + Q_{ij}^2}\} \quad (14)$$

where $V_i, V_j$ are the voltage magnitudes of the sending and receiving buses, respectively. The transmission line is modelled as a PI configuration, and $Z_{ij}\angle\theta_{ij}$ is its impedance and $b_{ij}$ is the susceptance.

The current flows $I_{ij}$ in each line is computed as (15).

$$I_{ij} = \frac{V_i\angle\delta_i - V_j\angle\delta_j}{Z_{ij}\angle\theta_{ij}} + \frac{b_{ij}V_i}{2}\angle(\delta_i + \frac{\pi}{2}) \quad (15)$$

Then the line losses $P_{l,ij}$ can be computed as (17).

$$P_{l,ij} = \text{real}\{I_{ij}^2 Z_{ij}\angle\theta_{ij}\} \quad (16)$$

The power generation for each generator $P_{gi}$ at bus *i* can be determined through its power balance between the demand $P_{Di}$ and power flow as (17).

$$P_{g,i} = \sum_{j \in \Omega_l^i} P_{ij} + P_{D,i} \\ Q_{g,i} = \sum_{j \in \Omega_l^i} Q_{ij} + Q_{D,i}\} \quad (17)$$

### B. Constraints

The generators and lines have constraints on their power rating as illustrated by (18) and (19) respectively. The bus voltage variation range (20) is also limited by grid code requirements.

$$X_{ST} = [P_{ST,i}, Q_{ST,i}, I_{drec,i}, I_{qrec,i}, I_{dinv,i}, I_{qinv,i}, m_{drec,i}, m_{qrec,i}, m_{dinv,i}, m_{qinv,i}, V_{s,i}]^T \quad (11)$$

$$P_{g,i}^{min} \leq P_{g,i} \leq P_{g,i}^{max} \atop Q_{g,i}^{min} \leq Q_{g,i} \leq Q_{g,i}^{max}\} \quad (18)$$

$$S_{ij}^{min} \leq S_{ij} \leq S_{ij}^{max} \quad (19)$$

$$V_i^{min} \leq V_i \leq V_i^{max} \quad (20)$$

Equation (14-17) and constraints (18-20) define the static power flow for the transmission system. The parameters are the line information $(Z_{ij}, \theta_{ij}, b_{ij})$. The variables are the generation at each bus $(P_{gi}, Q_{gi})$, current in each line $(I_{ij})$, the bus voltage $(V_i, \delta_i)$. Here, the matrix for the transmission system variables is defined as $X_S$ (21), and the constraints are defined as $X_S^{max}$(22), $X_S^{min}$ (23).

$$X_S = [P_{gi}, Q_{gi}, I_{ij}, V_i, \delta_i]^T \quad (21)$$

$$X_S^{max} = [P_{g,i}^{max}, Q_{g,i}^{max}, S_{ij}^{max}, V_i^{max}]^T \quad (22)$$

$$X_S^{max} = [P_{g,i}^{min}, Q_{g,i}^{min}, S_{ij}^{min}, V_i^{min}]^T \quad (23)$$

### C. System Model

The connection of the ST and the transmission system is at the PQ bus or the PCC point. The input from the transmission system to the ST is the bus voltage $V_i\angle\delta_i$, while the ST output is the demand $P_{STi}, Q_{STi}$ at bus $i$. However, here we assume that not every load is connected to the transmission system through an ST. Thus, here we introduce the judgement parameter matrix $G_{STi}$ and $G_{Li}$, the elements of which are defined as follows:

1) $G_{STi} = 1$ and $G_{Li} = 0$, if the load at bus $i$ is connected through a ST;

2) $G_{STi} = 0$ and $G_{Li} = 1$, if the load at bus $i$ is connected through a conventional low frequency transformer;

Consequently, the demand can be computed as (24).

$$P_{D,i} = G_{ST,i}P_{ST,i} + G_{L,i}P_{L,i} \atop Q_{D,i} = G_{ST,i}Q_{ST,i} + G_{L,i}Q_{L,i}\} \quad (24)$$

Note, in (24), the demand $(P_{D,i}, Q_{D,i})$ and PCC power $(P_{ST,i}, Q_{ST,i})$ are the variables, while the other terms are the parameters.

The complete system variable matrix and its constraints are as follows:

$$X = [X_S, X_{ST}, P_{D,i}, Q_{D,i}, P_{ST,i}, Q_{ST,i}]^T \quad (25)$$

$$X^{max} = [X_S^{max}, X_{ST}^{max}]^T \quad (26)$$

$$X^{min} = [X_S^{min}, X_{ST}^{min}]^T \quad (27)$$

### D. Objective Functioncs

The paper aims to address the system level benefit of including the ST in the economic dispatch, thus, the objective function is to minimize the total generation costs as (28).

$$MinF = \sum_{i=1}^{N}(a_{g,i}P_{g,i}^2 + b_{g,i}P_{g,i} + c_{g,i}) \quad (28)$$

$$X^{max} \leq X \leq X^{max} \quad (29)$$

## IV. CASE STUDY

The OPF formulation of the system is solved using the nonlinear programming (NLP) solver Conopt in GAMS. The test system is the New England IEEE 39-bus system as shown in Fig. 2. The generation bidding data is given in Table I. The base power is 100 MW. The ST capacity is set as 110% of the apparent power of the original load data. The resistance values $(R_{frec,i}, R_{DCh,i}, R_{DCm,i}, R_{finv,i})$ are picked in order to an ST efficiency of 96.5% efficiency at full load [23].

Table I. Generation costs (€/MW)

|  | G1 | G2 | G3 | G4 | G5 | G6 | G7 | G8 | G9 | G10 |
|---|---|---|---|---|---|---|---|---|---|---|
| $a_{g,i}$ | 0.0131 | 0.0111 | 0.0098 | 0.0071 | 0.0079 | 0.0213 | 0.0173 | 0.021 | 0.0013 | 0.0173 |
| $b_{g,i}$ | 13.32 | 13.32 | 20.7 | 20.93 | 21 | 10.52 | 5.47 | 5.47 | 10.52 | 10.52 |
| $c_{g,i}$ | 100 | 50 | 50 | 80 | 30 | 200 | 150 | 80 | 200 | 210 |

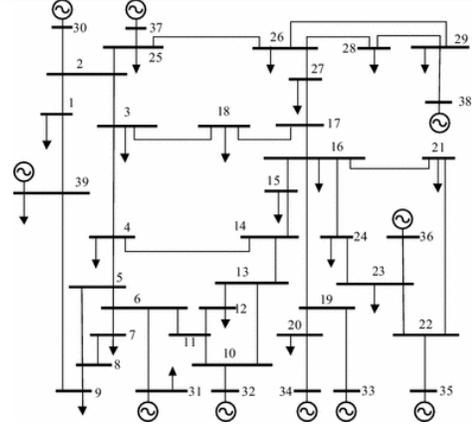

Fig. 2. New England IEEE 39-bus system

### A. System state comparison with no ST and full ST

In this case, the aim is to verify the effects of STs on the system states, i.e. $P_{Di}, Q_{Di}, V_i, P_{gi}$ by the comparison between a system with no ST, i.e. ($\{i| \forall G_{STi} = 0, G_{Li} = 1\}$) against a system with an ST at all demand interfaces ($\{i| \forall G_{STi} = 1, G_{Li} = 0\}$). To simplify the comparison, the system is operated at a full load situation, i.e. using the original loading data, and the voltage sensitivity is $\alpha = \beta = 1$, and the minimum demand voltage for all STs $V_{s,i}^{min}$ is 0.9 pu.

Fig. 3 shows comparison results for the active and reactive powers $P_{Di}, Q_{Di}$ at the transmission system buses. Note that in the ST case, the reactive powers on the primary and secondary sides are decoupled so that the reactive power required by the load is different from that at the transmission side, which has been used to optimise power flow. It can be seen from Fig. 3 (a) that the active demand for the full ST case for all the buses is lower than that in the no ST case. This is because the ST secondary voltage is independent of its primary voltage, so that for the sake of lower cost, the ST secondary voltage hits the lower limit, 0.9 pu, to reduce the active demand to 90% of its original value, referring to (5) with $\alpha = 1$. Although this is not a particularly realistic case as it makes no allowance for line drop and different load voltage sensitivities, it serves as an illustration of the ST operation. In addition, the reactive power at each bus is different between these two cases. This is because the reactive power in the ST is decoupled and can now vary from being inductive to capacitive in order to support the voltage and further optimize the power flow.

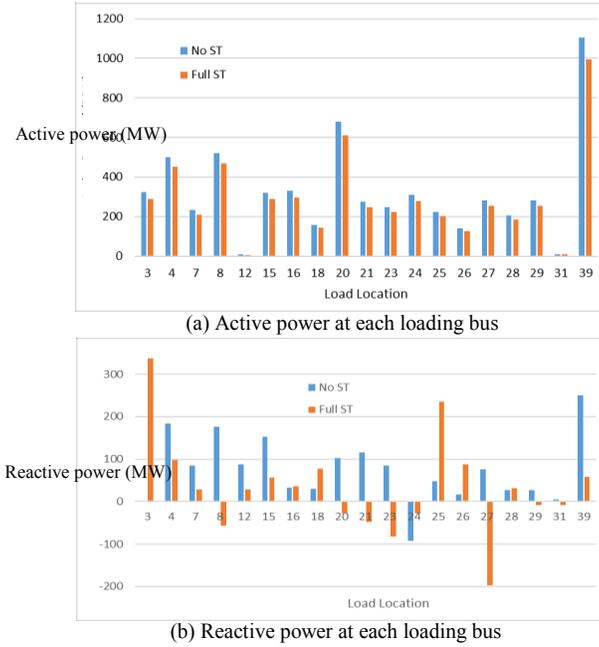

(a) Active power at each loading bus

(b) Reactive power at each loading bus

Fig. 3. No ST case vs. Full ST case on power comparison in the transmission system PCC point

Fig. 4 compares the bus voltages for all buses, where the x-axis represents the bus number. It can be observed that the bus voltages for the full ST case are (on average) higher than for the no ST case. This is due to reactive power compensation from the ST to raise the bus voltage and achieve lower line losses and further lower the generation cost.

Fig. 5 compares the generation from each generator. It can be seen that the demand reduction from the ST control generally reduce the generation only from those expensive generators. The losses can be obtained by the summation of the generation minus the summation of the demand. Here, the generation requirements for no ST is 6195 MW and the demand is 6149 MW, so that the losses are 46 MW. The generation requirement for the full ST case is 5574 MW and the demand is 5534 MW, so that the losses are 40 MW, again verifying that the higher voltages the lower losses.

### B. Vaying ST Penetration

In order to present a more realistic scenario, parameters such as loading and load voltage sensitivity should vary with time. In this case, we use a demand daily profile as shown in Fig. 6 to represent demand variation and analyse the effects of varying ST penetration on OPF benefits. Note, the parameters and limitations matrices $X_t^{max}, X_t^{min}$ now vary with time and correspondingly, the variable $X_t$ is time variant. The ST penetration is defined as the summation of the demand controlled by an ST over the total demand. The resolution is one hour. The case starts from no ST (0% penetration) in the system to full ST (100% penetration). We switch the directly connected load to the ST interfaced load gradually and record the total generation daily costs in this progress (see Fig. 7).

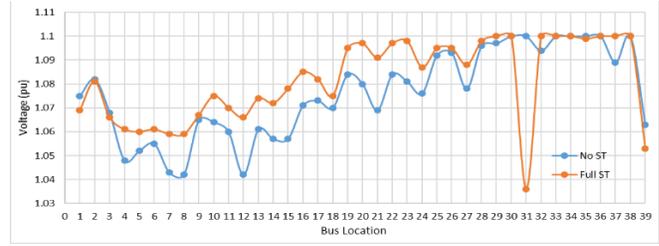

Fig. 4. No ST case vs. Full ST case on bus voltage comparison

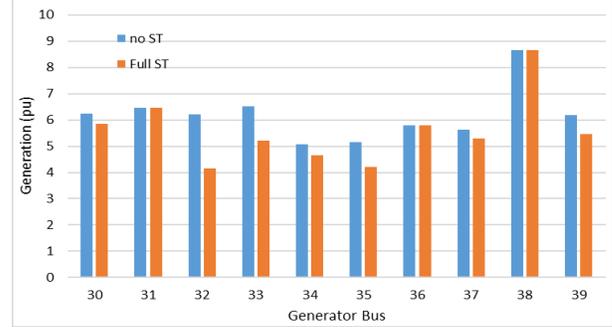

Fig. 5. No ST case vs. Full ST case on generation comparison

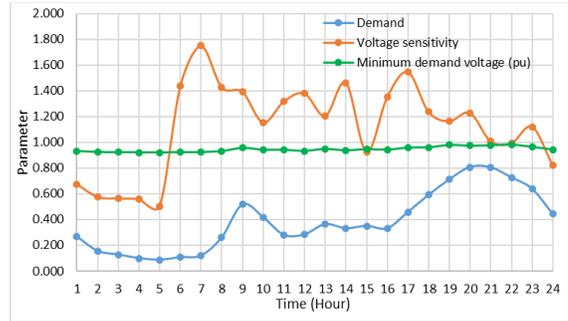

Fig. 6. Demand daily profile

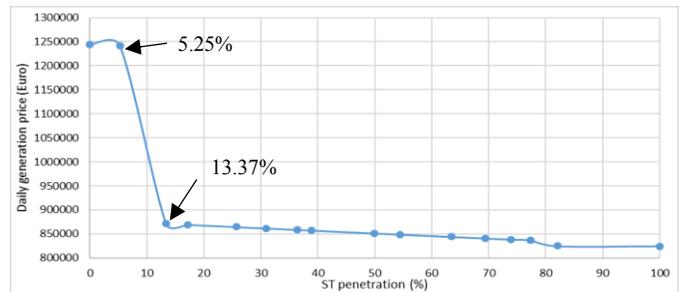

Fig. 7. The total generation cost versus ST penetration increase

It can be seen from Fig. 7 that the increase in ST penetration leads to a reduction on the total generation cost. It can be seen that between the 5.24% to 13.37% ST penetration case, there is a step reduction in the total generation daily costs. To illustrate the reason for this, the hourly generation costs are presented in Fig. 8 for these two scenarios along with the no ST and full ST scenario. It can be seen that the 5.25% ST penetration hourly costs are similar to the no ST scenario, and 13.37% ST penetration hourly costs is only slightly higher than the 100% ST scenario. However, there is an apparent gap between the 5.25% and 13.37% scenario. This is because in latter scenario, there is a generator downed off in the same period and the basic costs for this generator is removed.

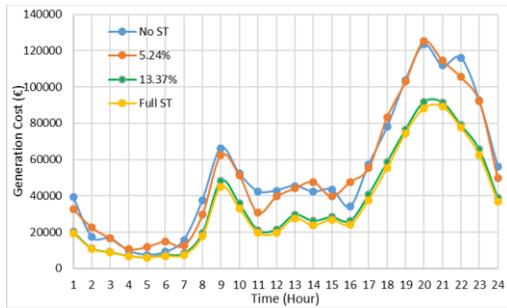

Fig. 8. Hourly generation costs.

## V. CONCLUSION

This paper has presented an ST model suitable for implementation in system level OPF analysis with the aim of evaluating the benefits of the ST as a controllable interface between the transmission and distribution systems. In particular the model captures the ST capability to independently provide reactive power to the transmission system and control demand through the utilisation of load voltage sensitivity. The model has been applied to a simple case study using the IEEE 39 bus system, which has shown some obvious advantages of independent reactive power provision and demand control in terms of reduced generation costs. The results from the investigation of increased ST penetration indicate that locational aspects may be of importance and that the location of STs at certain strategic locations may be particularly beneficial. Although in the analysis presented here, the load is simply reduced to reduce generation costs, and no account has been taken of the impact of this on customer energy use. Future work will investigate this aspect and also the possibility that the demand reduction capability be better utilised as fast frequency support and primary reserve.